\documentclass[preprint,12pt]{elsarticle}

\usepackage{amssymb,amsmath}
\usepackage{color}

\journal{Journal of Computational Physics}

\begin{document}

\begin{frontmatter}


\title{An Extended Krylov Subspace Model-Order Reduction Technique to Simulate Wave Propagation in Unbounded Domains}
\author{Vladimir Druskin}
\ead{Druskin1@slb.com}
\address{Schlumberger Doll Research, 1 Hampshire St., Cambridge, MA 02139, USA}
\author{Rob Remis}
\ead{R.F.Remis@TUDelft.NL}
\address{Circuits and Systems Group, Faculty of Electrical Engineering, Mathematics and Computer
Science, Delft University of Technology, Mekelweg 4, 2628 CD Delft, The Netherlands}
\author{Mikhail Zaslavsky}
\ead{mzaslavsky@slb.com}
\address{Schlumberger Doll Research, 1 Hampshire St., Cambridge, MA 02139, USA}
%

\title{An Extended Krylov Subspace Model-Order Reduction Technique to Simulate Wave Propagation in Unbounded Domains}



\begin{abstract}
In this paper we present a  novel extended Krylov subspace reduced-order modeling technique to efficiently simulate 
time- and frequency-domain wavefields in open complex structures. To simulate the extension to infinity, we use an optimal 
complex-scaling method which is equivalent to an optimized perfectly matched layer in which the frequency is fixed. Wavefields propagating 
in strongly inhomogeneous open domains can now be modeled as a non-entire function of the complex-scaled wave operator. Since this function contains 
a square root singularity, we apply an extended Krylov subspace technique to construct fast converging reduced-order models. Specifically, we use a modified version of the 
extended Krylov subspace algorithm as proposed by Jagels and Reichel [\emph{Linear Algebra Appl.}, Vol.~434, pp.~1716 -- 1732, 2011], since this algorithm allows us to balance the  
computational costs associated with computing powers of the wave operator and its inverse.  Numerical experiments from electromagnetics and acoustics illustrate the performance of the method. 
\end{abstract}


\end{frontmatter}


\section{Introduction}
\label{sec:intro}
Frequency- and time-domain wave problems in unbounded domains emerge in a wide range of applications: from seismic processing in geophysics to metamaterial development, from bioelectromagnetics to radar and antenna design. Furthermore, developing fast and robust forward modeling methods is not only a significant topic by itself, it is also of great importance in inversion and optimization algorithms. In all these different disciplines, conventional forward solution techniques are usually the method of choice. For time-domain problems, it is typically the Finite-Difference Time-Domain method (FDTD method, see, for example, \cite{Taflove&Hagness}) that is used, while discretized frequency-domain problems are usually solved using linear system solvers. When solutions for multiple frequencies are required, conventional workflow consists of solving frequency-domain problems for each frequency separately. In many applications, such methods quite often leave much room for improvement. Indeed, discretization grids for 3D problems may consist of up to a billion nodes and, even with state-of-the-art preconditioners, obtaining the solution of discretized frequency-domain problems results in rather computationally intensive tasks. 

Projection-based model reduction is a well-established tool and allows us to obtain time- and frequency-domain solutions by projecting the large scale dynamic system on a small Krylov or rational Krylov subspace. In an FDTD setting, for example, optimal time steps can be determined, though rather implicitly, via polynomial Krylov subspace projection methods. These Krylov methods have been shown to provide significant speed up for lossy diffusion-dominated problems, e.g.,  for the diffusive Maxwell equations (see \cite{ZDKGeophysics}). For lossless problems, however, e.g. wave problems in lossless media on bounded domains, such an optimality hardly provides any acceleration compared with time-stepping at the Courant limit \cite{DrKn89}.   

The principal difference between wave problems in bounded and unbounded domains is that the latter one is essentially lossy. Indeed, loosely speaking, infinity acts like an absorber and for any bounded subdomain there is no wave permanently traveling inside it without energy loss. Therefore, an attempt to construct a Krylov subspace projection method to speed up the solution of such problems is not hopeless. In fact, two of the authors showed that for wavefield problems on unbounded domains, standard polynomial Krylov methods may indeed outperform FDTD on large time intervals provided we respect the delicate spectral properties of the corresponding continuous problem (see \cite{Druskin&Remis}). To be more precise, in any practical implementation the domain of computation is necessarily bounded and boundary conditions have to be imposed for domain truncation. If selfadjoint boundary conditions (Dirichlet boundary conditions, for example) are used then the solution to a selfadjoint problem with a continuous spectrum is approximated by a solution to a selfadjoint problem with a discrete spectrum. In the latter case, the solution consists of an infinite summation of periodic eigenfunctions that do not decay. Enlarging the domain of computation does not help, since the eigenfunctions remain periodic and do not show any attenuation. For general sources then, polynomial Krylov methods used on discretized bounded domains with selfadjoint boundary conditions cannot outperform FDTD running at the Courant limit. However, for non-selfadjoint boundary conditions that simulate the extension to infinity, polynomial Krylov methods may outperform FDTD on large time intervals. This can be explained by considering the scattering poles expansion of a wavefield $u$ satisfying the wave equation on an unbounded domain~\cite{Lax&Phillips}. In particular, Lax and Phillips showed that for any fixed point in $\mathbb{R}^n$, $n$ odd, the solution $u(t)$ can be expanded for large $t$ in a sum of time-exponential modes. These modes decay exponentially in time and correspond to the so-called scattering resonances. The resonances are the poles of the resolvent for the wave equation and are located on the second Riemann sheet of the square root function. In practice, only a finite number of the corresponding modes essentially contribute to the solution for large $t$. Consequently, polynomial Krylov methods may outperform FDTD for large times if we can capture these resonant poles via a Krylov subspace method.    

The scattering poles can be found by truncating the domain of interest via complex coordinate scaling also known as Aguilar-Balslev-Combes-Simon theory~\cite{Hislop&Sigal}. This coordinate stretching technique was introduced in the 1970s and is since then used extensively in atomic and molecular physics to calculate energies, widths, and cross-sections of open quantum systems. The technique coincides with Berenger's perfectly matched layer (PML, \cite{Berenger}) using a fixed frequency for coordinate stretching. The problem with complex scaling is, however, that resonances and anti-resonances of the system are found simultaneously resulting in instability of the traditionally computed solution via a $\sin$ or $\cos$ function of the operator. Two of the authors resolved this issue in \cite{Druskin&Remis} by introducing a so-called stability-corrected exponential function that depends on the square root of the wave operator. Any type of Krylov subspace method can now be applied and we split them into two groups depending on the application. The first group targets to applications where the offline cost of constructing the subspace is more important compared to the online cost of subsequent projection. For example, as a stand-alone forward solver, the latter cost is typically negligible. \textit{Polynomial} Krylov subspace methods can be quite competitive here in terms of computation time, especially when the frequency range doesn't include small frequencies.  However, when applied within an inversion/optimization framework, not only the forward problem solution is required but also the derivatives with respect to the parameters that are being optimized or inverted. Depending on the number of these parameters, the computation of the derivatives can make the projection cost dominating. For these applications, multiple solutions of the problem are required, and the same reduced order model can be reused for this purpose.  In this case, it is beneficial to construct a subspace of smaller size for a given accuracy even at a larger offline cost. We put the methods of this kind into the second group. Taking this into account, we note that for non-entire operator functions (such as the stability-corrected operator function considered here), \textit{polynomial} Krylov subspace methods are known to be not the best choice in terms of convergence speed. Indeed,  \textit{rational} Krylov subspace (RKS) methods are known to be much better suited for non-entire functions (see, e.~g., \cite{Druskin&Knizhnerman}). Moreover, it has recently been shown that RKSs are beneficial for the solution of regular enough wave problems on bounded domains \cite{Grimm}. 

In this paper, we follow the approach based on the stability-corrected exponential function and construct optimal rational field approximants via a so-called Extended Krylov Subspace Method (EKSM). This Krylov method generates field approximations using both positive and negative matrix powers and produces rational field approximants with restricted poles at the origin and at infinity. The original EKSM of \cite{Druskin&Knizhnerman} allowed for short recursions and a structured projection matrix only for a fixed number of positive or negative matrix powers. More flexible EKSM modifications, allowing for sequences of nested extended subspaces with increasing positive and negative matrix powers, were suggested in \cite{Simoncini, JagelsReichel2009}. Here we use a modified version of the algorithm proposed in \cite{JagelsReichel2009}, since it allows for an arbitrary ratio between positive and negative matrix powers. We illustrate the performance of our proposed solution technique in Section~\ref{sec:numres}, where we present a number of numerical experiments related to the solution of the wave equation and solutions of Helmholtz's equation for a wide range of frequencies. In our experiments for time-domain problems, we show that our solution technique significantly outperforms polynomial Krylov subspace projection in terms of subspace dimension, while for Helmholtz's equation we show that with the PKS method we obtain solutions for a complete frequency range at the cost of an unpreconditioned linear solver applied to the problem with the smallest frequency. Our results clearly show that EKS significantly outperforms a polynomial Krylov approach when solutions for small frequencies are required.

\section{Problem formulation and polynomial Krylov subspace projection}

Let us consider the hyperbolic initial value problem 
\begin{equation}
\label{eq:cont}
A u + u_{tt} =0 
\qquad 
\text{with $\left.u\right|_{t=0}=0$ and $\left.u_{t}\right|_{t=0}=b$}
\end{equation}
for $t\geq 0$. Its frequency-domain counterpart is 
\begin{equation}
\label{eq:contfd}
A \mathsf{u} + s^2 \mathsf{u} =b.
\end{equation}
In these equations, $A$ is a selfadjoint nonnegative partial differential equation operator on an unbounded domain having an absolutely continuous spectrum. The solution of Eq.~(\ref{eq:cont}) can be written in terms of operator functions as 
\begin{equation}
\label{eq:sol_cont}
u(t) = \eta(t)A^{-1/2} \sin(A^{1/2} t) b,
\end{equation}
where $\eta(t)$ is the Heaviside unit step function. In \cite{Druskin&Remis, DrGutKni} we discretized the above partial differential equation on a uniform second-order grid and used Zolotarev's optimal rational approximation of the square root to implement an optimal complex scaling method for domain truncation. (The optimal scaling approach of \cite{Druskin&Remis} deals with propagative modes only, while in \cite{DrGutKni} a more advanced implementation is presented in which evanescent and propagative modes are taken into account simultaneously.) Writing the resulting system as  
\begin{equation}
\label{eq:discr}
\tilde{A}_{N} u_{N} + u_{N;tt} =0 
\qquad 
\text{with $\left.u_{N}\right|_{t=0}=0$ and $\left.u_{N;t}\right|_{t=0}=b_{N}$},
\end{equation}
where the tilde indicates that complex-stretching has been applied and the subscript~$N$ signifies that we are dealing with a discretized system, it is tempting to take the solution 
\begin{equation}
\label{eq:unstab_discr}
u_{N}(t) = \eta(t)\tilde{A}_{N}^{-1/2} \sin(\tilde{A}_{N}^{1/2} t) b_{N} 
\end{equation}
as an approximation to the exact solution as given by Eq.~(\ref{eq:sol_cont}). 
The approximation of Eq.~(\ref{eq:unstab_discr}) is unstable, however, since the eigenvalues of $\tilde{A}_{N}$ approximate both the resonances and the anti-resonances of the system. Therefore $\sin(\tilde{A}_{N}^{1/2} t)$ increases exponentially as $t$ increases and $u_{N}$ is unstable. 

Similarly, important frequency-domain field properties are lost if we take 
\begin{equation}
\label{eq:unstab_discr_fd}
\mathsf{u}_{N}(s) = 
\left(\tilde{A}_{N}+s^2I\right)^{-1}b_N
\end{equation}
as an approximation to $\mathsf{u}(s)$. In particular, the symmetry property $\mathsf{u}(\bar{s})=\bar{\mathsf{u}}(s)$ is broken and the limit $\lim_{s\rightarrow 0}{\mathsf{u}_{N}(s)}$ is not purely real as it should be, of course.

Fortunately, we can correct for these failures by exploiting the symmetry of the exact solution with respect to $(-\tilde{A}_{N})^{1/2}$ and $\overline{(-\tilde{A}_{N})^{1/2}}$, where the overbar denotes complex conjugation. Specifically, setting
\begin{equation}
\label{eq:defB}
B_{N} = 
\left(
-\tilde{A}_{N}
\right)^{1/2}
\end{equation}
and taking the principle value for the square root, a stabilized approximation to the continuous open problem of Eq.~(\ref{eq:cont}) is given by \cite{Druskin&Remis}
 \begin{equation}
 \label{eq:stabcor_time}
 \tilde{u}_{N}(t) = 
- \eta(t)\text{Re}
 \left[
B_{N}^{-1} \exp
 \left(
 -B_{N} t
 \right)
 \right]
 b_{N}.
 \end{equation}
This approximation is unconditionally stable because of the principle value convention. We note that if we replace $\tilde{A}_{N}$ by $A$ in Eq.~(\ref{eq:stabcor_time}), then the expression for the stability-corrected field becomes identical to the operator function expression for $u$ in Eq.~(\ref{eq:sol_cont}). In addition, by applying a Laplace transform to Eq.~(\ref{eq:stabcor_time}), we obtain the frequency-domain result 
\begin{equation}
\label{eq:stabcor_freq}
\tilde{\mathsf{u}}_{N}(s) =
-\frac{1}{2}
\left[
B_{N}^{-1}
\left(
B_{N} + s I_{N}
\right)^{-1}
+
\overline{B}_{N}^{-1}
\left(
\overline{B}_{N}+sI_{N}
\right)^{-1}
\right]
b_{N},
\end{equation}
where $I_{N}$ is the identity matrix of order~$N$. This stability-corrected field satisfies the symmetry property  $\mathsf{\tilde{u}}_{N}(\bar{s})=\bar{\mathsf{\tilde{u}}}_{N}(s)$ and becomes purely real as $s \rightarrow 0$. 
Finally, we note that for frequency-domain wave fields operating at an angular frequency $\omega$ we take $s=\text{i}\omega$, of course.

Having the stability-corrected field approximations available, we can now construct polynomial reduced-order models in the usual way. In particular, the models are drawn from the Krylov subspace 
\begin{equation}
\label{eq:pol_K}
\mathbb{K}_{m} = 
\text{span} 
\{
b_{N}, \tilde{A}_{N}b_{N},\tilde{A}_{N}^2 b_{N},...,\tilde{A}_{N}^{m-1}b_{N}
\}
\end{equation}
and are obtained by following a Petrov-Galerkin procedure. We remark that a basis for $\mathbb{K}_{m}$ can be generated very efficiently via a Lanczos-type algorithm using a short three-term recurrence relation, since matrix~$\tilde{A}_{N}$ is essentially complex-symmetric (see Section~\ref{sec:algorithm}). Furthermore, the error of the reduced-order models based on the Krylov space of Eq.~(\ref{eq:pol_K}) is of the same order as the error of the PML discretization in $\tilde A_N$ on the frequency interval of the signal. Finally, we note that matrix~$\tilde{A}_{N}$ is only required to form matrix-vector products in this algorithm and the cost per iteration is close to the cost of a single FDTD iteration. Further details can be found in \cite{Druskin&Remis}, where it is also shown that the reduced-order models generated in this way eventually do indeed outperform FDTD on large time intervals. 

\section{Extended Krylov Subspace Reduced-Order Models}
\label{sec:EKS_ROM}

Given that the stability-corrected field approximations of Eqs.~(\ref{eq:stabcor_time}) and (\ref{eq:stabcor_freq}) have a square root singularity, we observe that the reduced-order models taken from the Krylov subspace of Eq.~(\ref{eq:pol_K}) may not provide us with the fastest convergence. The reason is that models taken from this space are polynomial field approximations, but rational functions are usually much better at approximating functions near singularities (see \cite{Trefethen}).  This then naturally leads us to consider rational Krylov subspaces for stability-corrected field approximations. However, the main drawback of utilizing rational Krylov spaces is that large shifted systems of equations (Helmholtz systems) need to be solved to generate bases for these spaces. Without any effective preconditioners, this may simply be unfeasible. 
 
 To handle the square root singularity, we therefore resort to the extended Krylov subspaces
 \begin{equation}
 \mathbb{K}_{m_{1},m_{2}} = 
 \text{span}
 \{
 \tilde{A}_{N}^{-m_{1}+1}b_{N}, ...,\tilde{A}_{N}^{-1}b_{N}, b_{N}, \tilde{A}_{N}b_{N}..., 
 \tilde{A}_{N}^{m_{2}-1}b_{N}
  \}.
 \end{equation}
 An extended Krylov subspace can be seen as a special case of a rational Krylov subspace with one expansion point at zero and one at infinity. Obviously, the action of $\tilde{A}_{N}^{-1}$ on a vector is required to generate a basis for such a space. This essentially amounts to solving a Poisson-type equation for which efficient solution techniques are available~\cite{Golub&VanLoan}. Computing matrix-vector products with the inverse of the system matrix is generally still more expensive than computing matrix-vector products with the system matrix itself, however,  and from a computational point of view we therefore prefer to deal with extended Krylov subspaces~$\mathbb{K}_{m_{1},m_{2}}$ with $m_{1}<m_{2}$. In addition, the basis vectors should be constructed via short term recurrence relations, since storage of all basis vectors is generally not practical for large-scale real-world problems. 
 
In the original EKS method as introduced for Hermitian matrices in \cite{Druskin&Knizhnerman}, the number of inverse powers~$m_{1}$ is fixed and nested subspaces 
\[
\mathbb{K}_{m_{1},1} \subset \mathbb{K}_{m_{1},2} \subset ... \subset \mathbb{K}_{m_{1},m_{2}}
\]
are recursively generated via short term recurrence relations. Furthermore, it is shown that for the matrix square root and $m_{2}=m_{1}$, the number of iterations required for convergence of the EKS reduced-order models is of $O(\sqrt[4]{\| A_{N}\| \| A_{N}^{-1} \|})$. We stress that this estimate holds for Hermitian matrices~$A_{N}$ and diagonal ($m_{1}=m_{2}$) subspaces only. Generating a basis for such spaces obviously requires an equal number of iterations with the system matrix and its inverse. 

In the EKS method proposed in \cite{Simoncini}, again a diagonal extended Krylov subspace is considered, but this time the sequence of nested subspaces 
\[
\mathbb{K}_{1,1} \subset \mathbb{K}_{2,2} \subset ... \subset \mathbb{K}_{m_{1},m_{1}}
\]
is generated via short recurrence relations. The method converges at a rate as predicted by theory, but its computational costs are very large.   

In \cite{JagelsReichel2009, JagelsReichel2011}, Jagels and Reichel developed a more generalized approach in which the sequence of subspaces 
\[
\mathbb{K}_{1,i+1} \subset \mathbb{K}_{2,2i+1} \subset ... \subset \mathbb{K}_{k,ki+1}
\]
is generated again via short term recurrences. Here, $i$ is an integer that allows us to optimize the computational costs. In particular, since matrix-vector products with the system matrix are usually more easily computed than matrix-vector products with its inverse, it may be advantageous to apply the extended Krylov approach of Jagels and Reichel, since the number of iterations with the system matrix is essentially $i$ times larger than the number of iterations~$k$ that are carried out with the inverse of the system matrix. Different values of $i$ do produce different extended Krylov subspace sequences, however, and the reduced-order models based on these spaces may have different approximating properties. Furthermore, the Krylov projection matrix is a pentadiagonal matrix of order $k(i+1)$ with a block-tridiagonal structure and consequently the EKS reduced-order models can be evaluated very efficiently. 

In this paper, we use a variant of the extended Krylov scheme as proposed by Jagels and Reichel to construct EKS reduced-order models for wavefield problems in open domains. We exploit the complex symmetric structure of the system matrix to generate a basis of the extended Krylov space $\mathbb{K}_{k,ki+1}$. The basis vectors are complex orthogonal by construction and can be computed via short term recurrence relations. This scheme is presented in Section~\ref{sec:algorithm} and provides us with an extended Krylov decomposition of the system matrix~$\tilde{A}_{N}$. The computed decomposition then allows us to construct the EKS reduced-order models for the stability-corrected field approximation of Eqs.~(\ref{eq:stabcor_time}) and (\ref{eq:stabcor_freq}). 

 \section{Extended Krylov Subspace Algorithm} 
 \label{sec:algorithm}
As mentioned above, we are interested in constructing EKS reduced-order models for wavefield quantities in unbounded domains that satisfy the wave equation with variable coefficients. To be specific, let us consider wavefields propagating in two- or three-dimensional unbounded domains. In the interior, we discretize the wave equation on an equidistant second-order grid and use optimal Zolotarev complex-scaling to simulate the extension to infinity. We do not provide any details about this discretization process here, since it is completely described in \cite{Druskin&Remis,DrGutKni}. We do mention, however, that our approach is not limited to second-order schemes in the interior. Other higher-order discretization schemes can be used as well. In fact, since the optimal scaling-method has spectral accuracy, it would be preferable to use a discretization scheme that has spectral accuracy in the interior as well. High-order spectral methods can be applied for example (see \cite{Hesthaven_etal}) or optimal grids for interior domains as described in \cite{Asvadurov_etal}. Since our focus is on constructing EKS reduced-order models, we use a straightforward second-order accurate scheme for simplicity only.     

After the discretization procedure, we arrive at the semi-discretized system of Eq.~(\ref{eq:discr}).  Due to complex-scaling, the system matrix~$\tilde{A}_{N}$ belongs to $\mathbb{C}^{N \times N}$ and is not selfadjoint (Hermitian). It does, however, satisfy the symmetry relation
\begin{equation}
\tilde{A}_{N}^{T} M = M \tilde{A}_{N},
\end{equation}
where $M$ is a diagonal matrix with products of the spatial stepsizes and medium parameters on its diagonal (for details, see  \cite{Druskin&Remis}). Note that matrix~$M$ is not positive definite and does not even belong to $\mathbb{R}^{N \times N}$, since the optimal stepsizes used for complex-scaling are complex.

Introducing now the bilinear form 
 \begin{equation}
 \label{eq:bil}
 \langle x,y \rangle = y^{T} M x
 \end{equation}
we have $\langle \tilde{A}_{N} x,y \rangle = \langle x,\tilde{A}_{N} y \rangle$ for any $x,y \in \mathbb{C}^{N}$ and two (nonzero) vectors $x$ and $y$ belonging to $\mathbb{C}^{N}$ are said to be complex orthogonal if $\langle x,y \rangle =0$. Finally, we stress that Eq.~(\ref{eq:bil}) does not define an inner product, since $M$ is not positive definite.  

Given the symmetry of matrix~$\tilde{A}_{N}$, we can now generate a complex orthogonal basis for $\mathbb{K}_{k,ki+1}$ by modifying the orthogonalization algorithm for $\mathbb{K}_{k,ki+1}$ and real symmetric matrices of Jagels and Reichel~\cite{JagelsReichel2009, JagelsReichel2011}. In particular, if we replace the inner products for vectors from $\mathbb{R}^{N}$ by the bilinear form of Eq.~(\ref{eq:bil}), we arrive at an algorithm that generates a complex orthogonal basis of the Krylov space $\mathbb{K}_{k,ki+1}(\tilde{A}_{N},b_{N})$, where each basis vector has a Euclidean norm equal to one. We note that the algorithm may break down, since the bilinear form does not define a definite inner product. 
With $\| \cdot \|$ denoting the Euclidean norm, the algorithm is as follows: 
\newpage

\bigskip
\noindent
\textbf{Algorithm} Complex Orthogonalization Process for $\mathbb{K}_{k,ki+1}(\tilde{A}_{N},b_{N})$
\hrule
\begin{small}
         
\bigskip
\textbf{Input:} $k$, $i$, and source vector~$b_{N}$

\medskip
\textbf{Output:} complex orthogonal basis $\{ v_{j} \}_{j=-k}^{ki+1}$ of $\mathbb{K}_{k,ki+1}(\tilde{A}_{N},b_{N})$   

\bigskip
$v_{-1}=0$; 

$\beta_{0}=\| b_{N} \|$; $v_{0} = b_{N}/\beta_{0}$; $\delta_{0}=\langle v_0, v_0 \rangle$;

\medskip
\begin{itemize}
\item[] \textbf{for} $p=0,1,...,k-1$ \textbf{do}
\begin{itemize}
\item[] $u:= \tilde{A}_{N} v_{-p}$;
\item[] $\alpha_{-p,ip} := \langle u,  v_{ip} \rangle$; $u:=u - \alpha_{-p,ip}/\delta_{ip} v_{ip}$;
\item[] $\alpha_{-p,-p}:= \langle u,v_{-p} \rangle$; $u:=u-\alpha_{-p,-p}/\delta_{-p} v_{-p}$;
\item[] $\beta_{ip+1}:= \| u \|$; $v_{ip+1}:=u/\beta_{ip+1}$; $\delta_{ip+1}=\langle v_{ip+1}, v_{ip+1} \rangle$;

\medskip
\item[] $u:= \tilde{A}_{N} v_{ip+1}$;
\item[] $\alpha_{ip+1,ip} := \langle u,v_{ip} \rangle$; $u:=u-\alpha_{2p+1,ip}/\delta_{ip}v_{ip}$; 
\item[] $\alpha_{ip+1,-p} := \langle u,v_{-p} \rangle$; $u:=u-\alpha_{ip+1,-p}/\delta_{-p}v_{-p}$;
\item[] $\alpha_{ip+1,ip+1} := \langle u,v_{ip+1} \rangle$; $u:=u-\alpha_{ip+1,ip+1}/\delta_{ip+1}v_{ip+1}$;
\item[] $\beta_{ip+2}=\| u \|$; $v_{ip+2}=u/\beta_{ip+2}$; $\delta_{ip+2}=\langle v_{ip+2}, v_{ip+2} \rangle$;

\medskip
\item[] \textbf{for} $j=3,...,i$ \textbf{do}
\begin{itemize}
\item[] $u:= \tilde{A}_{N} v_{ip+j-1}$;
\item[] $\alpha_{ip+j-1,ip+j-2} := \langle u,v_{ip+j-2}\rangle$; $u:=u-\alpha_{ip+j-1,ip+j-2}/\delta_{ip+j-2} v_{ip+j-2}$;
\item[] $\alpha_{ip+j-1,ip+j-1} := \langle u,v_{ip+j-1}\rangle$; $u:=u-\alpha_{ip+j-1,ip+j-1}/\delta_{ip+j-1} v_{ip+j-1}$;
\item[] $\beta_{ip+j}:= \| u \|$; $v_{ip+j}=u/\beta_{ip+j}$; $\delta_{ip+j}=\langle v_{ip+j}, v_{ip+j} \rangle$;
\end{itemize}
\item[] \textbf{end}
\item[] $w:=\tilde{A}_{N}^{-1}v_{i(p+1)}$;
\item[] $\delta_{i(p+1),-p} := \langle w,v_{-p}\rangle$; $w:=w-\delta_{i(p+1),-p}/\delta_{-p}v_{-p}$;
\item[] \textbf{for} $j=1,...,i$ \textbf{do}
\begin{itemize}
\item[] $\delta_{i(p+1),ip+j} := \langle w, v_{ip+j} \rangle$; $w:=w-\delta_{i(p+1),ip+j}/\delta_{ip+j} v_{ip+j}$; 
\end{itemize}
\item[] \textbf{end}
\item[] $\beta_{-(p+1)}=\| w \|$; $v_{-(p+1)}=w/\beta_{-(p+1)}$; $\delta_{-p+1}=\langle v_{-p+1}, v_{-p+1} \rangle$;
\end{itemize}
\item[] \textbf{end}
\end{itemize}
\end{small}

\newpage
\noindent 
After a successful completion of this algorithm, we store the vectors of the generated basis in the $N$-by-$k(i+1)$ matrix 
 \begin{equation}
 \label{eq:matV}
 V_{k(i+1)}=[v_{0},v_{1},...,v_{i},v_{-1},...,v_{-k+1},...,v_{ik}]
 \end{equation}
 and all iterations can be summarized into the equation 
 \begin{equation}
 \label{eq:Krylov_decomp}
 \tilde{A}_N  V_{k(i+1)} =  V_{k(i+1)} H_{k(i+1)} + z_{k(i+1)} e_{k(i+1)}^{T},
 \end{equation}
where 
\[
z_{k(i+1)} = h_{k(i+1)+1,k(i+1)} v_{-k} + h_{k(i+1)+2,k(i+1)} v_{ik+1}
\]
and $h_{ij}$ is the $(i,j)$ entry of matrix~$H_{k(i+1)}$. Furthermore, matrix~$H_{k(i+1)}$ is a pentadiagonal matrix (see \cite{JagelsReichel2009}) that satisfies 
\begin{equation}
\label{eq:mat_H}
D_{k(i+1)}H_{k(i+1)} = V_{k(i+1)}^{T} \tilde{A}_{N} V_{k(i+1)},
\end{equation}
where $D_{k(i+1)}$ is diagonal matrix with entries $\delta_0,\delta_1,...,\delta_{i},\delta_{-1},...,\delta_{-k+1},...,\delta_{ik}$. The entries of matrix $H_{k(i+1)}$ can be easily obtained in explicit form from the ones given in \cite{JagelsReichel2009}.

Equation (\ref{eq:Krylov_decomp}) is the extended Krylov decomposition of the system matrix~$\tilde{A}_{N}$ and can be used to construct the EKS reduced-order models both in the time- and frequency-domain. Specifically, setting $d=k(i+1)$ and introducing
\begin{equation}
\label{eq:def_Bred}
B_{d} = 
\left(
-D_d^{1/2}H_{d}D_d^{-1/2}
\right)^{1/2},
\end{equation}
where again the principle value for the square root is taken, the time-domain EKS reduced-order model is given by
\begin{equation}
\label{eq:EKS_ROM_time}
\tilde{u}_{d}(t) = 
-\eta(t) \delta_{0}
\text{Re}
\left[
V_{d} B_{d}^{-1} \exp(-B_{d}t) 
\right]
e_{1},
\end{equation}
where $e_{1}$ is the first column of the identity matrix of order~$d$. Its frequency-domain counterpart is given by 
\begin{equation}
\label{eq:EKS_ROM_freq}
\tilde{\mathsf{u}}_{d}(s) =
-\frac{1}{2} \delta_0
\left[
V_{d} B_{d}^{-1}
\left(
B_{d} + s I_{d}
\right)^{-1}
+
\overline{V}_{d}
\overline{B}_{d}^{-1}
\left(
\overline{B}_{d}+sI_{d}
\right)^{-1}
\right]
e_{1}.
\end{equation}

 \section{Numerical Experiments} 
 \label{sec:numres}
To illustrate the performance of the EKS method in the time- and frequency-domain, two sets of experiments are presented. In the first set an acoustic wavefield problem in the frequency-domain is considered, while in the second set we consider an electromagnetic photonic waveguide problem in the time-domain.  

\begin{figure}[!htb]
  \centering
  \includegraphics[width=0.48\textwidth]{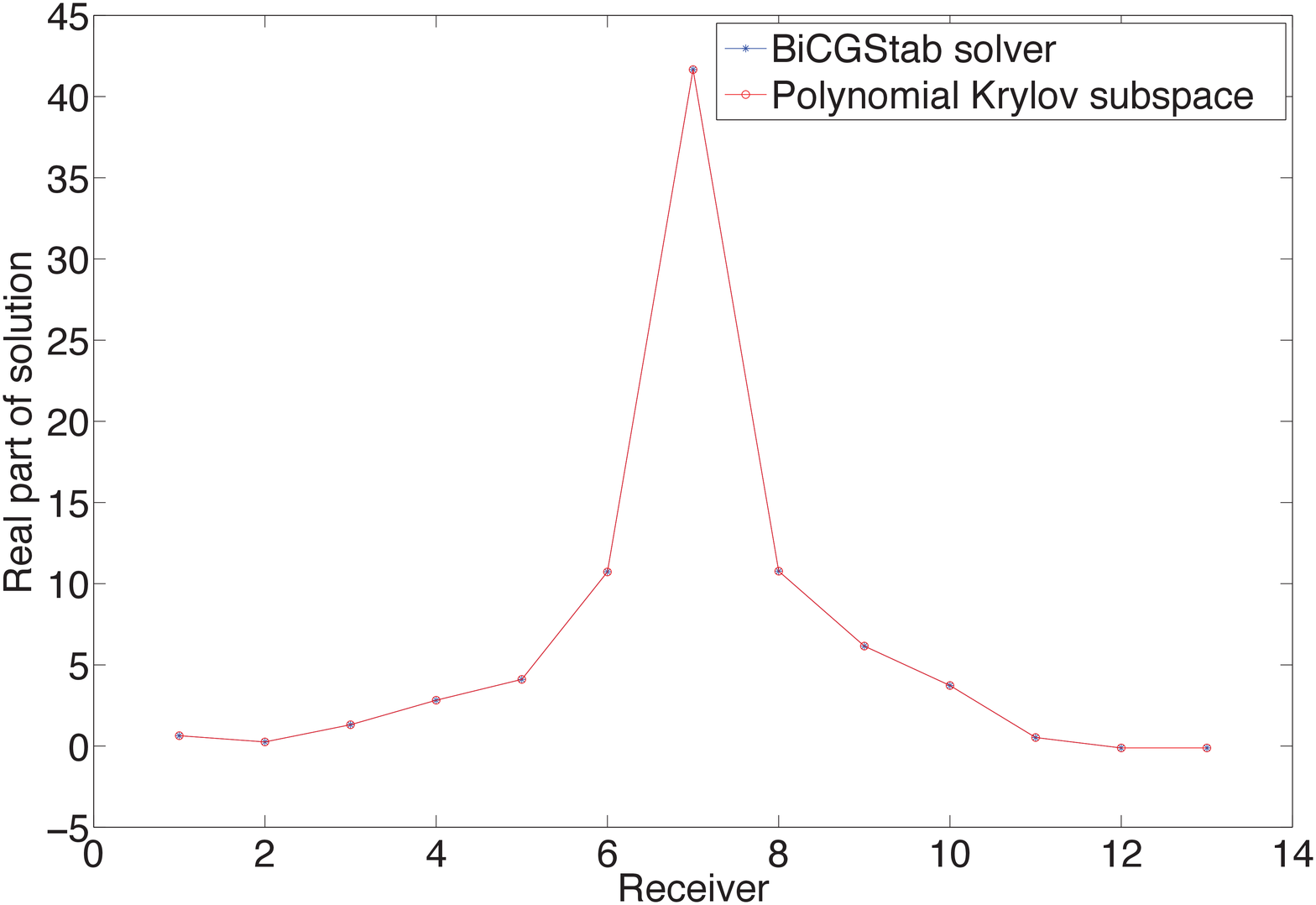}
  \includegraphics[width=0.48\textwidth]{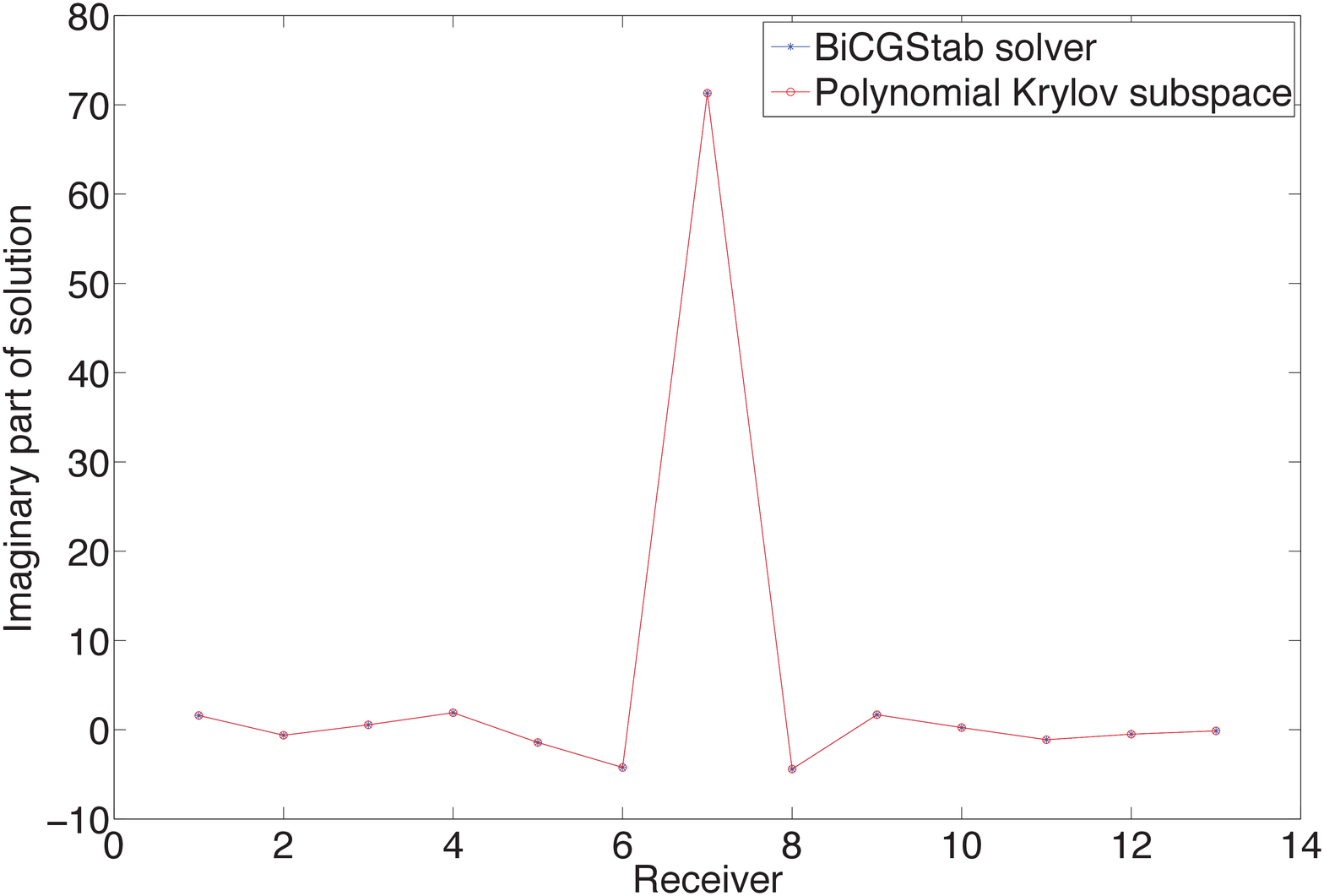} 
  \caption{SEG/EAGE Salt model. Real and imaginary parts of solution for frequency 7.5Hz computed using the BiCGStab and PKS solvers with 1600 matrix-vector multiplications and 1800 iterations,  respectively. Both solutions are almost indistinguishable.}
\label{fig:SEG1}
\end{figure}

\begin{figure}[!htb]
  \centering
  \includegraphics[width=0.48\textwidth, height=4.0cm]{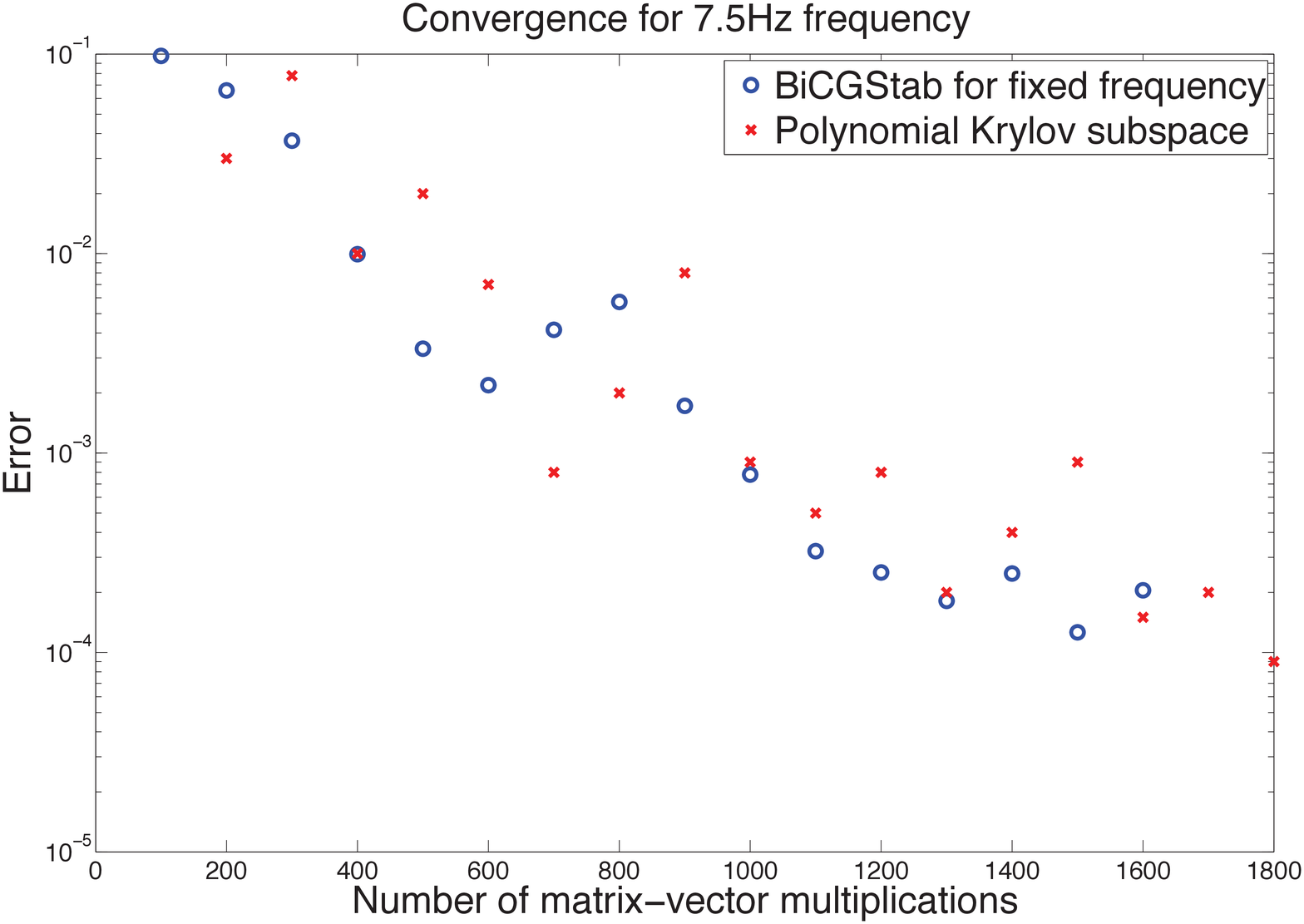}
  \includegraphics[width=0.48\textwidth, height=4.0cm]{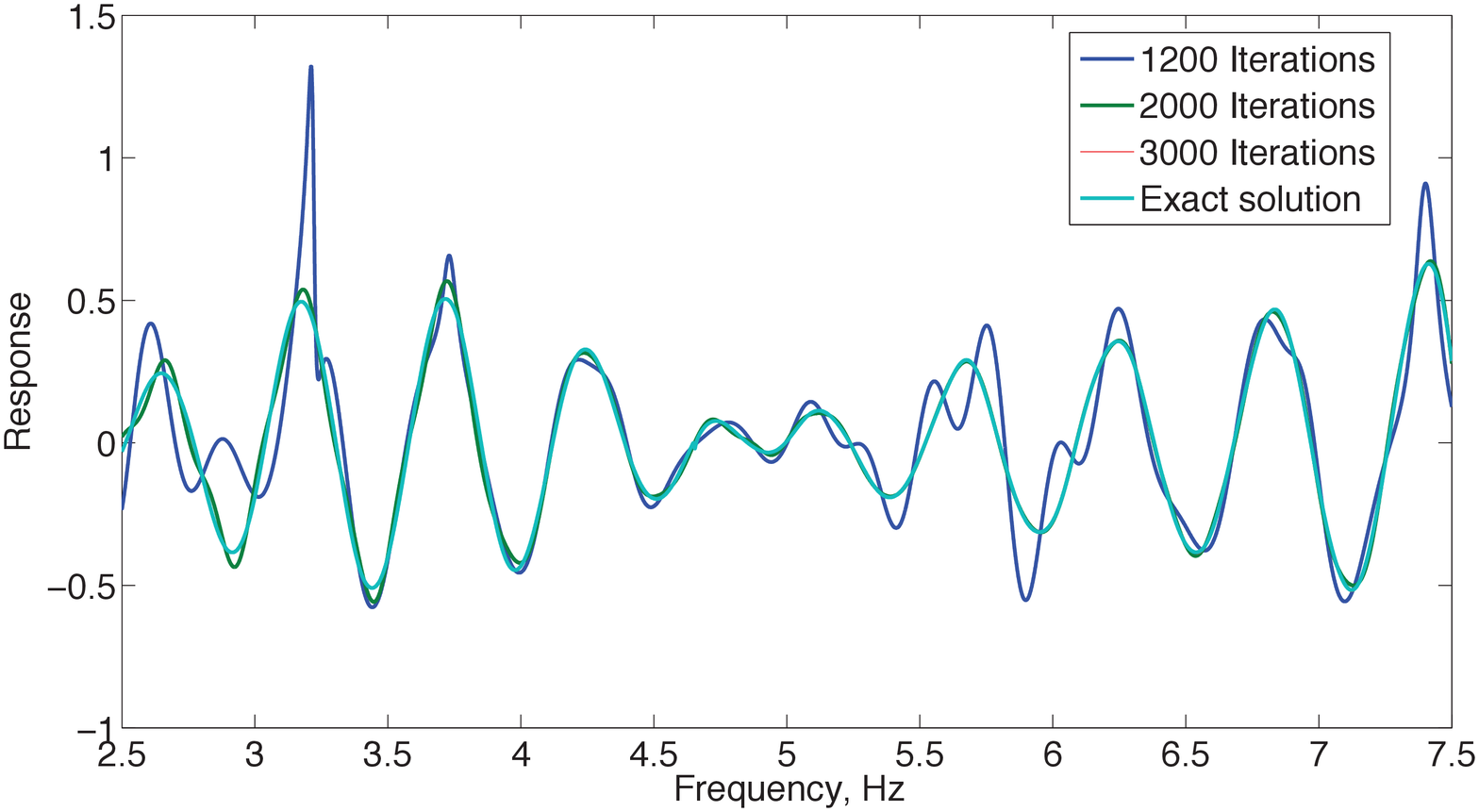}
  \caption{SEG/EAGE Salt model. Convergence rates for the problem with fixed frequency 7.5Hz (left) and convergence behaviour of PKS for frequency range 2.5Hz to 7.5Hz.}
\label{fig:SEG2}
\end{figure}

For the acoustic problem, we have considered a 3D SEG/Salt model and benchmarked our solver against an independently developed unpreconditioned BiCGStab algorithm that is  a competitive conventional iterative Helmholtz solver (see \cite{Pan_etal} for details). Here we took the case $m_1=1$ corresponding to PKS. Figure~\ref{fig:SEG1} shows the solutions at different receiver locations for a fixed frequency of 7.5Hz. Both the BiCGStab and PKS model reduction solutions coincide with each other very well. When comparing the number of iterations, we note that the PKS approach requires one matrix-vector multiplication per iteration, while BiCGStab needs two. Since this part constitutes the most computationally intensive part, it makes sense to compare the performance of the two methods in terms of matrix-vector multiplications rather than in terms of iterations. To this end, we have plotted the convergence rates of BiCGStab and model order reduction against the number of matrix-vector multiplications in Figure~\ref{fig:SEG2}~(left) again for a frequency of 7.5Hz.  We observe that for this single frequency problem both rates are rather close. We note that both approaches converge faster for higher frequencies and slower for lower ones. This is shown explicitly for the model reduction approach in Figure~\ref{fig:SEG2}~(right), where the convergence of Krylov approximations of a fixed order is shown for different frequencies within a frequency range running from 2.5Hz to 7.5Hz. At this point, we stress that the principal difference between the PKS method and BiCGStab is that the former approach obtains solutions for the whole given frequency range at the convergence cost of BiCGStab for the lowest frequency (from that range).

\begin{figure}[!htb]
  \centering
  \includegraphics[width=0.48\textwidth, height=4.0cm]{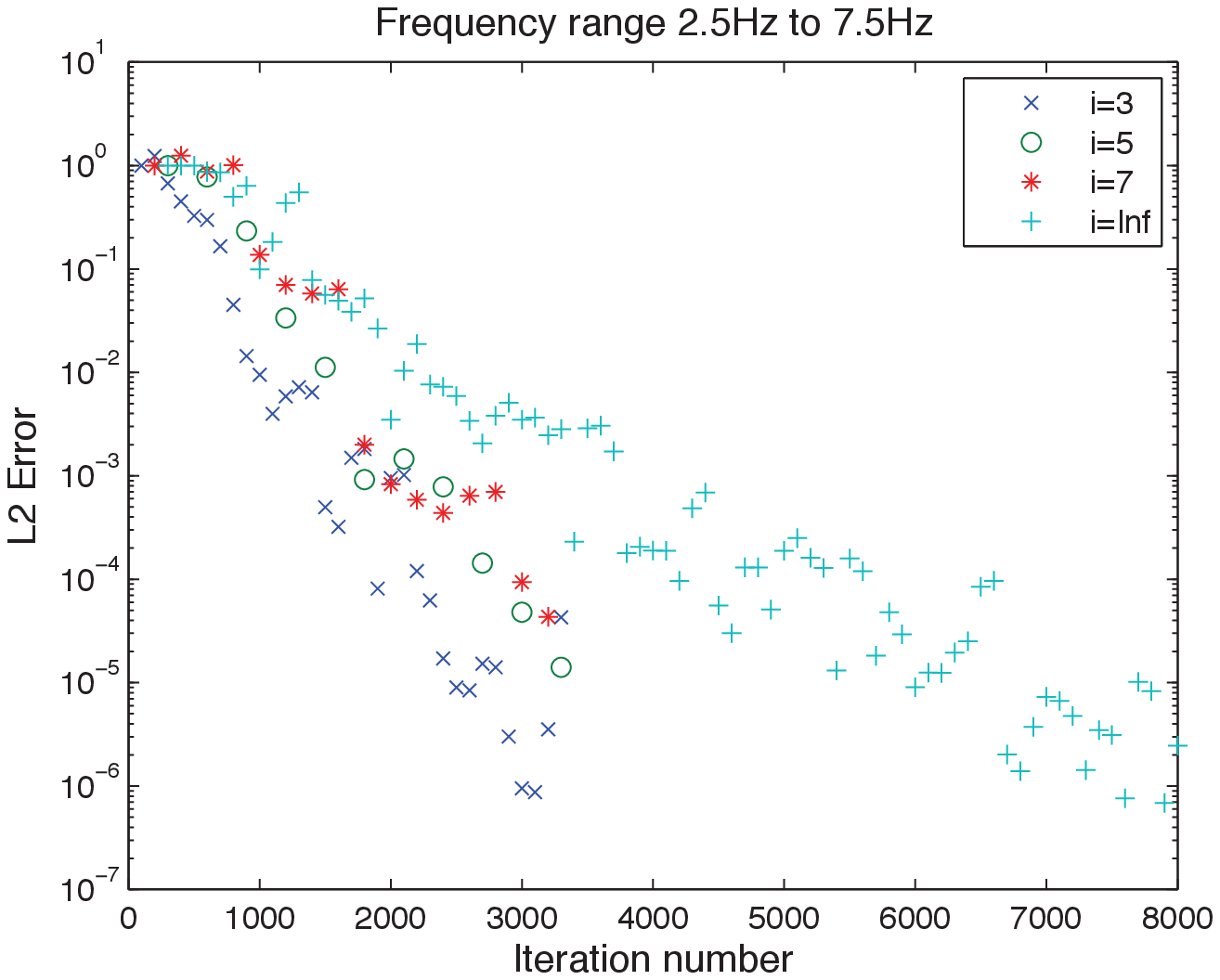}
  \includegraphics[width=0.48\textwidth, height=4.0cm]{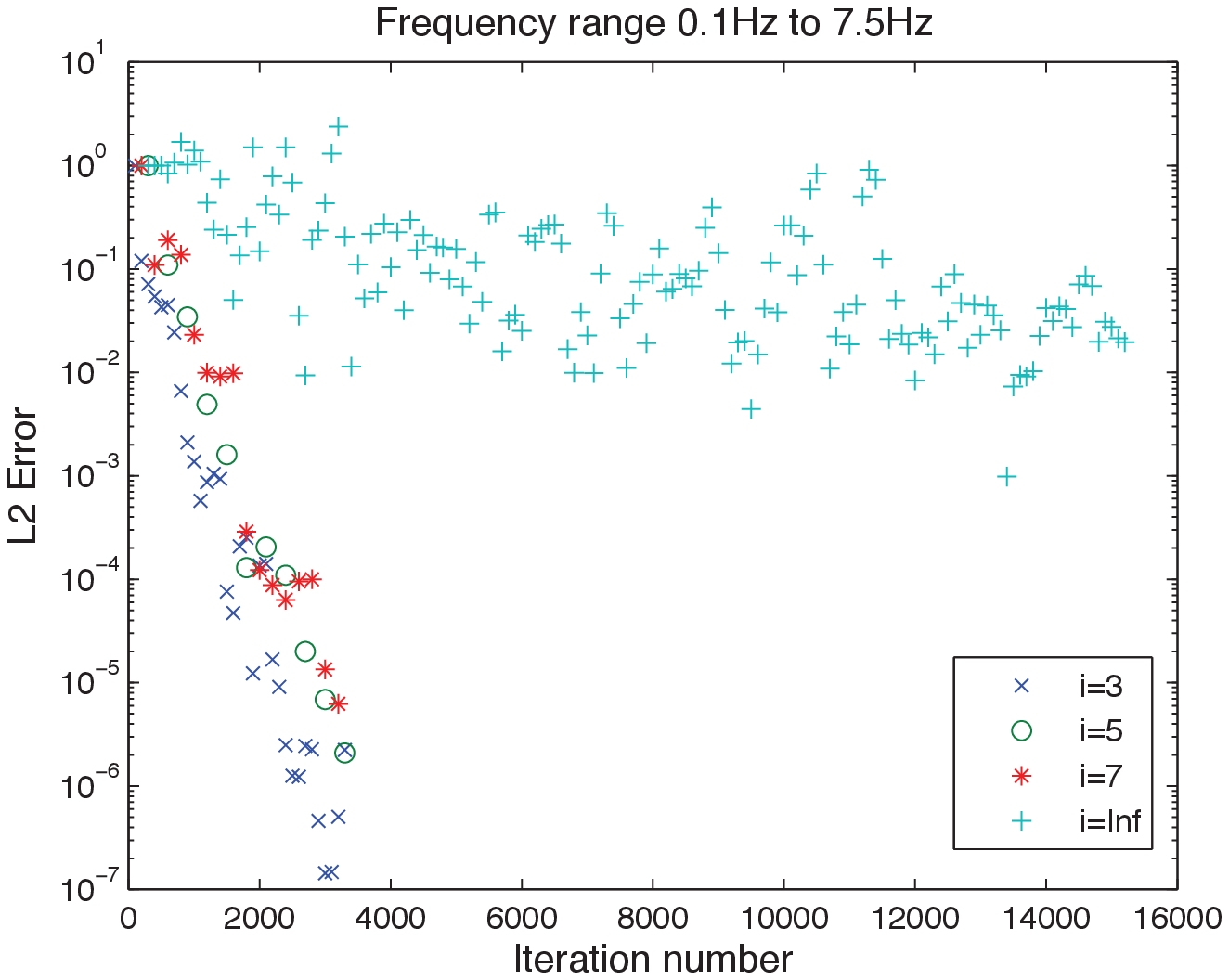}
  \caption{SEG/EAGE Salt model. Convergence of EKS for different values of $i$ and for frequency ranges from 2.5Hz to 7.5Hz (left) and from 0.1Hz to 7.5Hz (right).}
 \label{fig:SEG3}
\end{figure}

Subsequently, we consider what adding negative powers gives us in terms of performance. In Figure~\ref{fig:SEG3}~(left), we have plotted convergence curves (with respect to increasing $k$) for approximants obtained using EKS $\mathbb{K}_{k,ki+1}$ for different values of fixed $i$ ($i=\infty$ corresponds to PKS) and for a frequency range that runs from from 2.5Hz to 7.5Hz.  From this figure it is clear that adding negative powers ($i<\infty$) visibly improves EKS convergence compared with the PKS method. In terms of computation time, it took PKS 157500 seconds (4500 iterations) to reach $10^{-4}$ accuracy, while EKS with $i=3$, $i=5$, and $i=7$ required 154000 seconds (2300 iterations), 189000 seconds (3000 iterations), and 156000 seconds (300 iterations), respectively, to reach the same level of accuracy. 
From these experiments, it is clear that PKS converges faster for higher frequencies, while EKS improves convergence for smaller frequencies and, consequently, makes the convergence more uniform for the whole frequency range. This statement is confirmed in Figure~\ref{fig:SEG3}~(right), where we have plotted the convergence curves for the EKS method with the same subspace parameters as before, but this time for a frequency range that runs from 0.1Hz to 7.5Hz. We observe that PKS significantly slows down in this case, while EKS performs almost as efficient as for the narrower frequency range. For example, to reach a $10^{-4}$ accuracy, it takes about the same time for EKS to converge, while PKS hardly reached $10^{-2}$  accuracy in 525000 seconds of computational time (15000 iterations). 


\begin{figure}[t]
\begin{center}
\scalebox{0.45}{\includegraphics{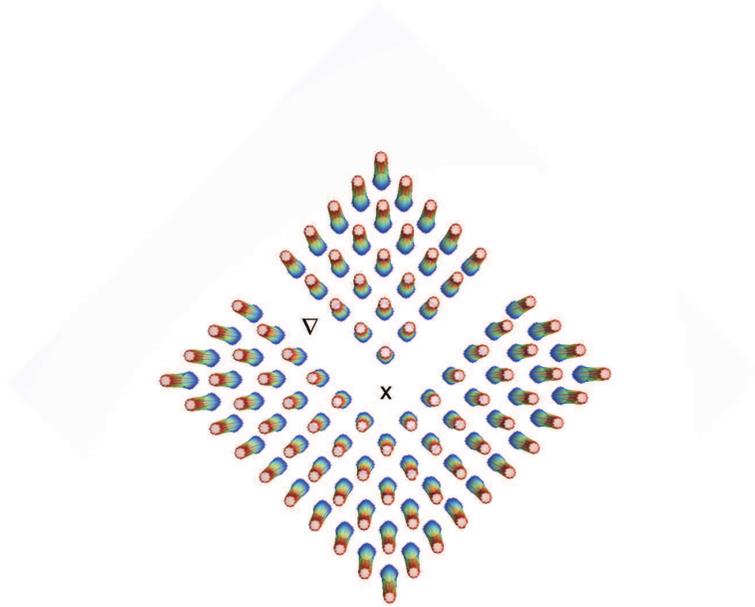}} 
\end{center}
\caption{Crystal consisting of dielectric rods. The cross and the triangle indicate the location of the source and receiver, respectively. The rods have a relative permittivity of 11.56. The distance~$a$ between the rods is 0.58~$\mu\text{m}$ and the radius of each rod is $0.18a$.}
\label{fig:crystal_conf}
\end{figure}

In our second set of experiments, we consider E-polarized time-domain electromagnetic wave field propagation in the two-dimensional configuration shown in Figure~\ref{fig:crystal_conf}. This configuration was also considered in \cite{Druskin&Remis}, where the PKS method was used to compute the field distributions. We revisit this problem here so that we can compare the performances of EKS and PKS. As an independent check, we also compare our results with FDTD computations. Since FDTD is based on the first-order Maxwell system, we use a first-order formulation of the EKS method in our time-domain experiments as well. 
 
The configuration of Figure~\ref{fig:crystal_conf} consists of a set of dielectric rods with vacuum as a background medium. The rods have a relative permittivity of 11.56 and the distance~$a$ between the rods is approximately 0.58~$\mu\text{m}$. The radius of each rod is $0.18a$. By removing rods inside this crystal we create a bend as is illustrated in Figure~\ref{fig:crystal_conf}. A source is placed at the corner of the bend and we record the electric field strength halfway one of the bends. The source pulse is a modulated Gaussian with its spectrum essentially contained in the frequency interval $9.8 \cdot 10^{14} \leq \omega \leq1.44 \cdot 10^{15}$~rad/s, which coincides with the ideal frequency bandgap of this structure.

\begin{figure}[t]
\begin{center}
\scalebox{0.3}{\includegraphics{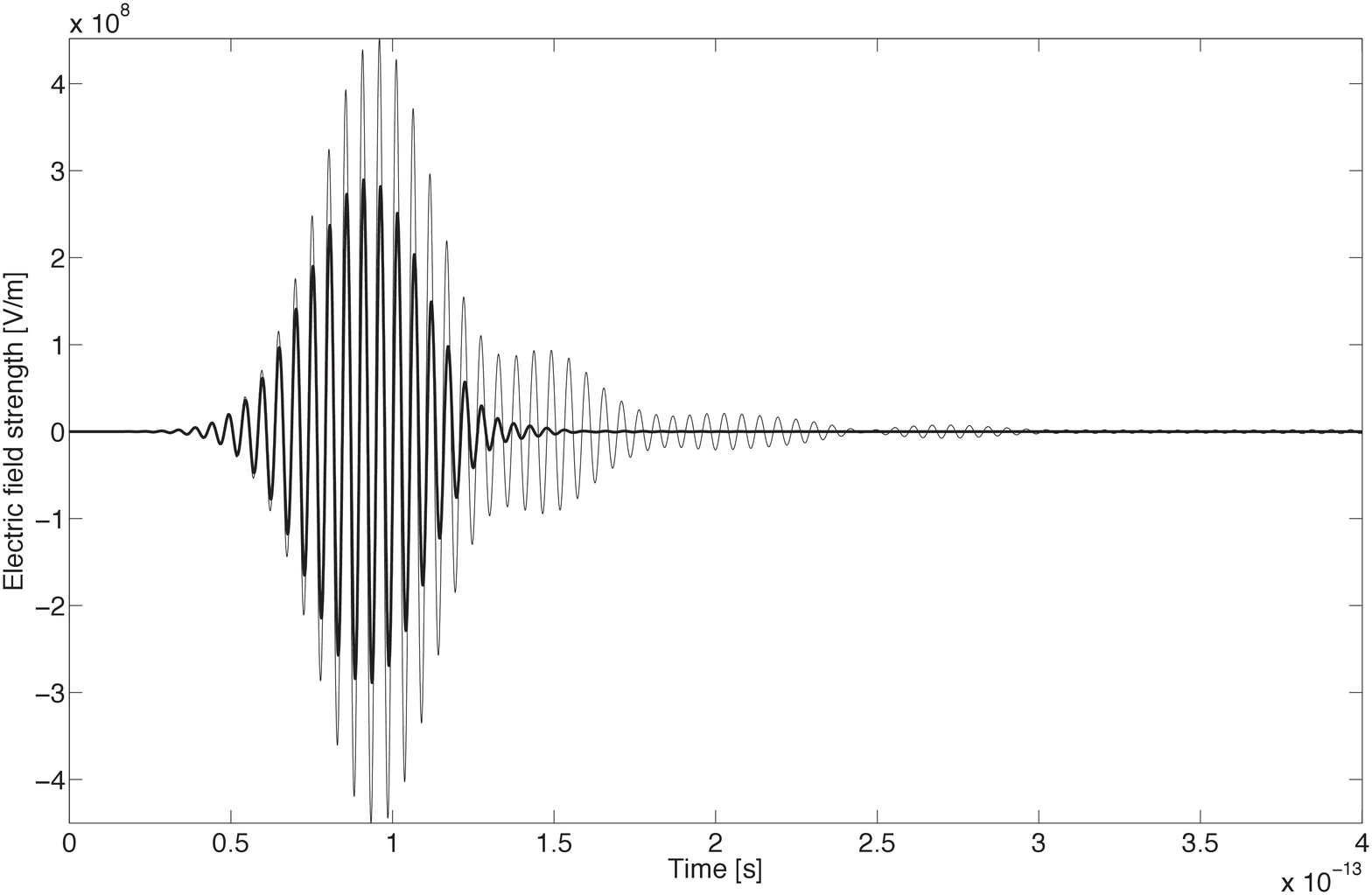}} 
\end{center}
\caption{Electric field strength at the receiver location. Black line: diagonal EKS reduced-order model after 100 iterations with the system matrix and 100 iterations with its inverse. Grey line: FDTD simulation obtained after 8197 iterations at the Courant limit.}
\label{fig:100it}
\end{figure} 

\begin{figure}
\begin{center}
\scalebox{0.3}{\includegraphics{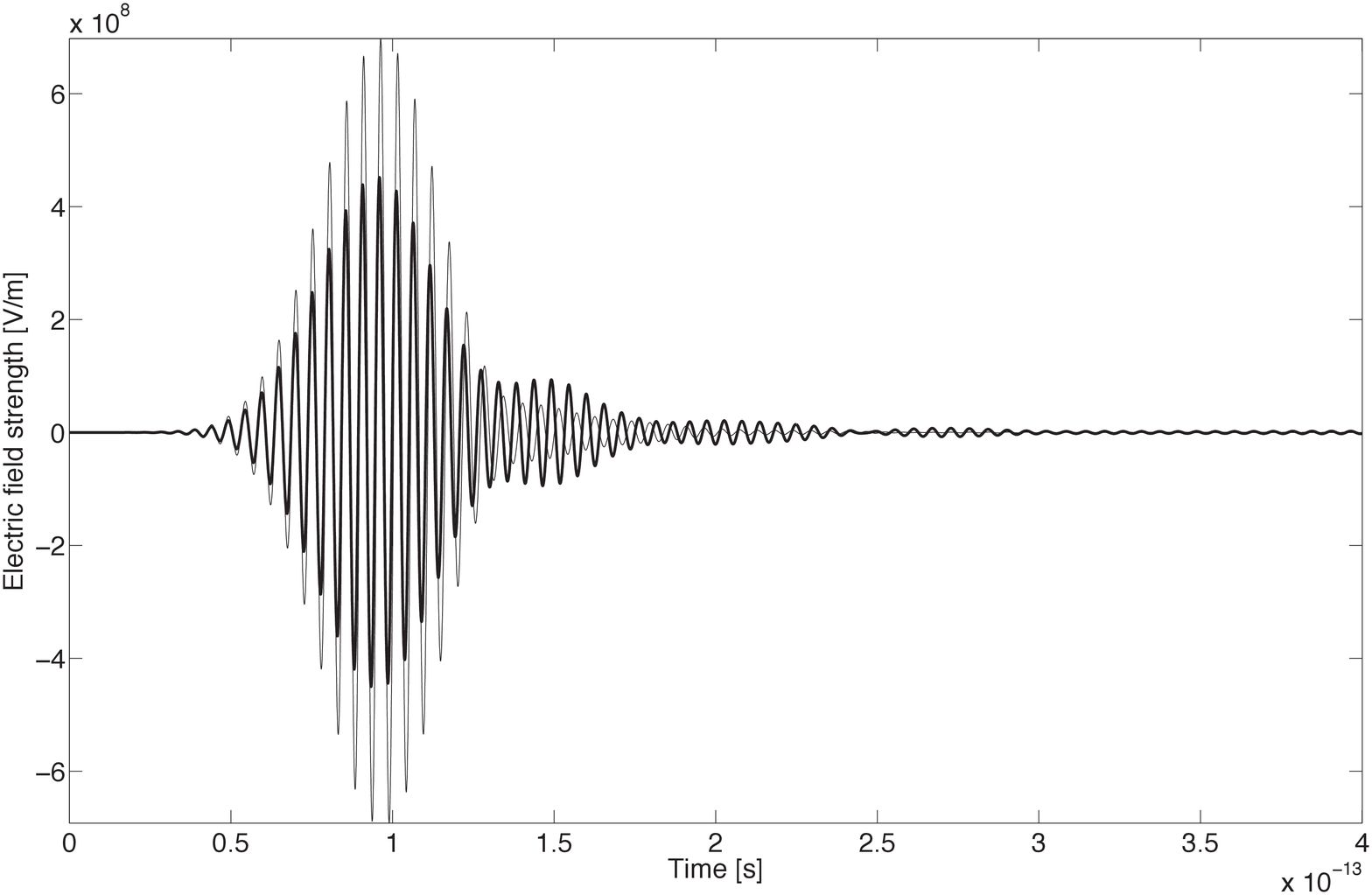}} 
\end{center}
\caption{Electric field strength at the receiver location. Black line: diagonal EKS reduced-order model after 200 iterations with the system matrix and 200 iterations with its inverse. Grey line: FDTD simulation obtained after 8197 iterations at the Courant limit. }
\label{fig:200it}
\end{figure} 

\begin{figure}
\begin{center}
\scalebox{0.3}{\includegraphics{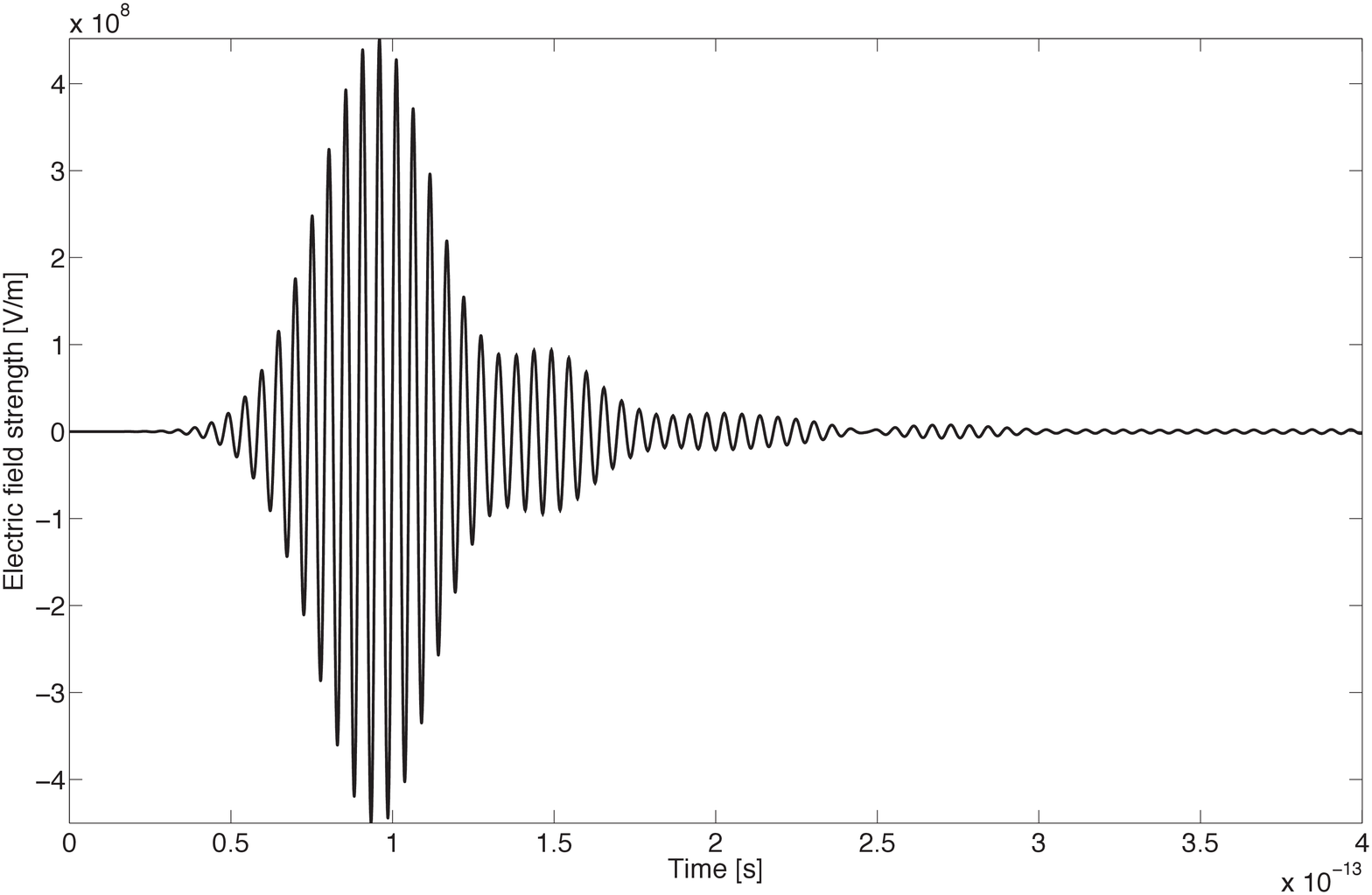}} 
\end{center}
\caption{Electric field strength at the receiver location. Black line: diagonal EKS reduced-order model after 400 iterations with the system matrix and 400 iterations with its inverse. Grey line: FDTD simulation obtained after 8197 iterations at the Courant limit.}
\label{fig:400it}
\end{figure}

In Figure~\ref{fig:100it}, we present the electric field response at the receiver location for a diagonal extended Krylov subspace ($m_{1}=m_{2}$) after 100 iterations with the system matrix and 100 iterations with its inverse (black line). The field response obtained after 8197 iterations of the FDTD method run at the Courant limit is also shown in Figure~\ref{fig:100it}. Clearly, the EKS reduced-order model has not converged yet on the time interval of interest. We therefore double the number of iterations and carry out 200 iterations with the system matrix and 200 with its inverse. The resulting EKS reduced-order model for the electric field strength is plotted in Figure~\ref{fig:200it} along with the corresponding FDTD result. Even though there is some improvement, the reduced-order model still has not converged on our time interval of observation. Finally, carrying out 400 iterations with the system matrix and 400 with its inverse we obtain the reduced-order model as shown in Figure~\ref{fig:400it} (black line). The model essentially overlaps with the FDTD result in this case and the number of iterations required by the EKS method is clearly much smaller than the number of FDTD iterations (800 for EKS, 8197 for FDTD). However, the amount of work that is involved at each iteration is not the same of course, since EKS requires matrix inversion, while FDTD does not. We therefore compare the computation times of both methods as well and find that FDTD requires 1764 seconds to solve this problem, while the EKS method  requires only 888 seconds. In terms of offline costs, we have achieved a speed up of approximately a factor of two compared with FDTD. If we compare this data with the results presented in \cite{Druskin&Remis}, we observe that EKS is actually slower than PKS in this case, since according to \cite{Druskin&Remis}, PKS requires only 606 seconds to solve this problem. However, the order of the PKS reduced-order model is 3000 (see \cite{Druskin&Remis}), which is much larger than the order of the EKS reduced-order model (which is 800). We conclude that for this problem both EKS and PKS are faster than FDTD. When reduction of the offline costs is important then PKS should be used, since it has the shortest computation time, but if the online costs are dominant then the EKS model should be used, since it provides us with the smallest reduced-order model. 

\section{Conclusions}
 \label{sec:conclusions}
Based on model reduction techniques, we have developed a powerful tool for solving large-scale multi-frequency wave problems. The PKS method allows us to obtain wave field solutions on a whole frequency range at the cost of an unpreconditioned linear solver applied to the problem with the smallest frequency. Moreover, we have shown that EKSs significantly outperform the polynomial approach when solutions for small frequencies are required and we have demonstrated that the EKS approach allows for substantial dimensionality reduction in the time-domain. The time-domain models produced by EKS have a much smaller order than the models produced by PKS. On the other hand, subspace creation via EKS takes more time than subspace creation via PKS. Based on these findings, we conclude that PKS should be used if reduction of the offline computational costs is desired, while EKS is the method of choice when the online costs need to be minimized. 

Finally we note that our approach is not limited to second-order schemes in the interior. In fact, since the optimal discrete PML has spectral accuracy (see \cite{Druskin&Remis, DrGutKni}), it would be preferable to use a discretization scheme that has spectral accuracy in the interior as well. High-order spectral methods can be applied, for example, (see \cite{Hesthaven_etal}) or optimal grids for interior domains as described in \cite{Asvadurov_etal}. 

\section{Acknowledgements}

We'd like to thank Dr. Guangdong Pan for providing us with results for his BiCGStab finite-difference solver.

\end{document}